\documentstyle[pre,aps,epsf]{revtex}
\begin{document}

\draft

\title{Level Statistics of Multispin-Coupling Models\\ with First and Second
Order Phase Transitions}

\author{Jean Christian Angl\`es d'Auriac}
\address{Centre de Recherches sur les Tr\`es Basses Temp\'eratures,
B.P. 166, F-38042 Grenoble, France}

\author{Ferenc Igl\'oi}
\address{Research Institute for Solid State Physics, 
H-1525 Budapest, P.O.Box 49, Hungary\\ Institute for Theoretical Physics,
Szeged University, H-6720 Szeged, Hungary\\ }

\date{January, 1998}

\maketitle

\begin{abstract}
\baselineskip=16pt
We consider self-dual transverse-field Ising spin chains with $m$-spin
interaction, where the phase transition is of second and first order,
for $m \le 3$ and $m>3$, respectively. We present a statistical analysis
of the spectra of the Hamiltonians on relatively large $L \le 18$ finite
lattices. Outside the critical point we found level repulsion close to
the Wigner distribution and the same rigidity as for the
Gaussian Orthogonal Ensemble. At the transition point the level statistics
in the self-dual sector is shown to be the superposition of two independent
Wigner distributions. This is explained by the existence of an extra
symmetry, which is connected to level crossing in the thermodynamic limit.
Our study has given no evidence for the possible integrability of the models
for $m>2$, even at the transition point.
\end{abstract}


\section{Introduction}
There are many problems in physics in which multi-particle interactions
play an important r\^ole. One may mention nuclear forces, solid $^3\rm He$
\cite{roger}, adsorbed systems\cite{benneman} and plasmas\cite{held}.
It is known from some exact results\cite{baxterwu} that the critical
properties of models with many body forces generally depend on
the range of interaction. Recently considerable effort has been made
to clarify the properties of a simple one-dimensional quantum model
described by the Hamiltonian\cite{turban,penson82}:
\begin{equation}
{\cal H} =-J\sum_l \sigma_l^x \sigma_{l+1}^x\dots \sigma_{l+m-1}^x-
h \sum_l \sigma_l^z\
\equiv -J H_x - h H_z \;.
\label{e:hamilton}
\end{equation}
Here the $\sigma_l^x$, $\sigma_l^z$ are Pauli matrices at site $l$ and
$J$ and $h$ are the exchange coupling and the transverse field,
respectively. The classical statistical mechanical equivalent of this
model is a two-dimensional square lattice Ising model with mixed
$m$-spin and two-spin interactions\cite{debierre83}.

The Hamiltonian Eq.~(\ref{e:hamilton})
 is self-dual\cite{turban,penson82} and the
self-dual point is $J=h$ independent of $m$. According to numerical
studies
\cite{penson82,debierre83,igloi83,igloi86,blote86,alcaraz86,kolbpenson,vanderzande,igloi87}
there is one phase transition in the
system, which takes place at the self-dual point, and the
transition changes from second to first order, when $m>m_c=3$. In the
borderline case $m=3$ the transition is
conjectured\cite{debierre83} to belong to the four state Potts
universality class, a conjecture which is supported by an approximate
mapping\cite{blote87} and by numerical
studies\cite{blote86,vanderzande,igloi87,alcaraz87}.

Concerning the simple structure of the model, its self-dual symmetry and
the expected relation to Q-state Potts models, one can also pose the
question, whether the model is integrable, at least in its self-dual
point. To find an answer to this question in this paper we are going to
study the
statistical properties of the spectrum of the Hamiltonian. As it has been
established in a series of papers
\cite{MoPoBeSi93,HsAdA93,PoZiBeMiMo93,BrAdA96,jc1,jc2,jc3} the spectrum of a
Hamiltonian (or the transfer matrix for classical statistical mechanical
models) has different statistical properties for integrable and
non-integrable models and one can make a close connection to the theory
of the spectral properties of random matrices. In the actual calculation
we first make use of all all those symmetries of the Hamiltonian which do not
depend on the value of the couplings, and for large finite lattices we
block-diagonalize the eigenvalue matrix of the problem. The statistical
analysis of the energy levels is then performed for each block separately.
In a non-integrable model, in which no further internal symmetry is
present the matrix-elements of a block-matrix are expected to be
loosely correlated, so that they can be approximately represented
by random entries. Indeed the spectrum of non-integrable models is found
to belong to the class of orthogonal random matrices, to the so called
Gaussian Orthogonal Ensemble (GOE) and the level spacing distribution is
described by the Wigner surmise\cite{wigner}:
\begin{equation}
P(s)={\pi \over 2} s \exp(-\pi s^2/4)\;.
\label{e:wigner}
\end{equation}
On the other hand if the Hamiltonian is integrable by the Bethe ansatz
there is an infinite number of internal symmetries and consequently
the matrix-elements of a block-matrix are strongly correlated. 
Loosely speaking integrable Hamiltonians are so peculiar
that they are not well described by an `average Hamiltonian'. Then one
expects that in this case the eigenvalues themselves behave like independent
random numbers, so that the spectrum of integrable models belongs to the
ensemble of diagonal random matrices and the level spacing distribution
is described by the Poissonian (exponential) distribution: $P(s)=\exp(-s)$.
Numerical studies of integrable models
\cite{MoPoBeSi93,HsAdA93,PoZiBeMiMo93,BrAdA96,jc1,jc2,jc3} are indeed
in agreement with this assumption.

In this paper we are going to perform the analysis of the level statistics
of the multispin coupling Hamiltonian in Eq.~(\ref{e:hamilton}). We are
going to answer two questions. The first question is
whether or not the Hamiltonian
is integrable, at least at the transition (self-dual) point. Our second
question concerns the characteristics of the level distribution at a first
order transition point. The paper is organized as follows. The symmetries
of the Hamiltonian in Eq.~(\ref{e:hamilton}), which are essential to perform
a block-diagonalization, are presented in Section 2. The statistical
analysis of the spectrum of the block-diagonalized Hamiltonian is given
in Section 3, while the results are discussed in the final Section.

\section{Symmetries of the Hamiltonian}

As described in the Introduction the first step in a statistical
analysis of the energy eigenvalues is to block-diagonalize the
Hamiltonian in Eq.~(\ref{e:hamilton}) using all those symmetries of
the model which do not depend on the actual values of the couplings.
Before analyzing these symmetries, let us first notice that
if $E(J,h)=\{E_0(J,h), E_1(J,h), \cdots,E_{2^L-1}(J,h)\}$ denotes
the set of energies of the Hamiltonian Eq.~(\ref{e:hamilton}),
one has $E(J,h) = E(\pm J,\pm h)$. This can be seen introducing
the operators $O^x = \prod_{i=0}^{L-1}{\sigma_i^x}$
and $O^z = \prod_{i=0}^{L/m-1}{\sigma_{i m}^z}$ and noting that
$H_\alpha O^\beta = \epsilon_{\alpha \beta} O^\beta H_\alpha$,
where $\alpha,\beta = x, z$ and $\epsilon_{\alpha,\beta}=1$ 
if $\alpha=\beta$ and -1 otherwise. On a finite lattice this symmetry
holds for periodic boundary conditions and if the length of the chain
is a multiple of $m$. In what follows we consider this type of lattices and
restrict ourself to the case $h>0$ and $J>0$.

The symmetries of the model are of three types: i) {\it space-like symmetries},
which describe invariance of the system under geometrical transformations
(translation, inversion, etc);~ii) {\it gauge symmetries}, which are
connected to invariance of the Hamiltonian under internal
transformations; and finally
iii) {\it duality symmetries}, which make a connection between the
strong- and weak-coupling regimes of the Hamiltonian.

i) The {\em space symmetry} of the model on a finite lattice depends on
the boundary condition. In a statistical analysis of the spectrum of
finite systems it is desired to use the most symmetric boundary condition
to get a block structure, which well represents
the statistical behavior of the spectrum in the thermodynamic limit. 
Therefore, as already mentioned, we apply periodic boundary conditions, 
which can be formally expressed as $\sigma_{L+i}^x=\sigma_i^x$. 
The space symmetry group is then the automorphy group of a ring, 
irrespective of the range of the interaction $m$. This is the
dihedral group generated by the translation $T$ and the reflection $R$,
($T^L=R^2={\rm Identity}$ and $T R= R T^{L-1}$) which both
commute with ${\cal H}$.

ii) The {\em gauge symmetries} are generalizations of the spin-reversal
symmetry for the well known $m=2$ case. Recalling that we 
take $L$ to be a multiple of $m$, let us introduce a set of $n=2^{m-1}$
operators $O_k$ for $k=0,1,\dots,n-1$:
\begin{equation}
O_k = \prod_{\alpha=0}^{L/m-1}
{\;\;\prod_{i=\alpha m}^{\alpha m+m-1}{(\sigma_i^z)^{k_i}}}
\end{equation}
where $k_0,k_1,\cdots,k_{m-2}$ are the bits of the binary representation
of $k$, and $k_{m-1}$ is such that 
\begin{equation}
\sum_{i=0}^{m-1}{k_i} \;\;{\rm even}\;. 
\label{e:cond}
\end{equation}
The spin reversal symmetry, which holds when $m$ is even,
is $O_{n-1}$ corresponding to $k_i=1$ for all $i$.
It is straightforward to check that
the condition Eq.~(\ref{e:cond}) ensures that 
all the operators  ${O_i}$ commute with ${\cal H}(J,h)$. 
It is also clear that these operators are diagonal, involutive
and form an Abelian group ($O_0$ is the identity).
This implies that all the $2^{m-1}=n$
representations are one-dimensional and the corresponding projectors
are of the form $P_R = 1/n \sum_{i=0}^{n-1}{\epsilon_i^R O_i}$
where $\epsilon_i^R =\pm 1$ for all $i$ and $R$.
All these projectors split the Hilbert space in $2^{m-1}$
invariant subspaces of size $2^{L-m+1}$ each.
For example, for $m=2$ and $L$ even, $P_0=(1/2)(O_0+O_1)$ projects
onto the subspace with an even number of up spins, while
$P_1=(1/2)(O_0 - O_1)$ projects onto the subspace with an odd
number of up spins.
The projector $P_0 = 1/n \sum{O_i}$ projects onto the most
symmetric subspace to which the ground state belongs
(we refer now to this subspace as the ground-state sector),
whereas the other $2^{m-1}-1$ sectors become degenerate in the
thermodynamic limit. Thus, in this limit, the
degeneracy of the ground state  in the strong coupling phase $J>h$ is 
given by $2^{m-1}$. This degeneracy for the $m=3$ model is just four,
which led Debierre and Turban\cite{debierre83} to conjecture the
same universality class for the transition as that for the $Q=4$ state
Potts model.

The combination of the space symmetry and the gauge symmetry
is not obvious, since the operators of these two groups
do {\em not} commute in general. The product of these
two groups is a semi-direct product, (not a direct product)
since the gauge group is a normal subgroup. As usual the states are labelled
by the number of the representation, $R$, to which they belong.
We have computed the character table of the symmetry group
from which the dimensions of the invariant subspaces are deduced and then the
block-diagonal Hamiltonian is constructed. The dimensions of the irreducible
representations and the size of the corresponding blocks are given in
Tables I-III for different values of $m$, in the range of $0\le R < L/2+3$ and
$0\le R < (L-1)/2+2$, for $L$ even and odd, respectively.
We note that in the ground state sector, which is labelled by $R=0$ 
and corresponds to 
$P_0=(1/n) \sum{O_i}$
all the operators of the space and gauge symmetry group commute,
so that {\em in this sector} we have a representation of the
dihedral group. 
In what follows we use the same labelling convention
as in \cite{jc2}.

iii) As mentioned in the Introduction the Hamiltonian in Eq.~(\ref{e:hamilton})
has the property of {\it duality symmetry}. To show this and its consequences
in finite lattices, first we define, 
for an infinite lattice, dual Pauli operators $\tau_i^x,~\tau_i^z$
as
\begin{eqnarray}
\label{e:dualpaulix}
\tau_i^z   &=& \sigma_i^x \sigma_{i+1}^x \dots \sigma_{i+m-1}^x~ \\
\sigma_i^z &=& \tau_i^x \tau_{i+1}^x \dots \tau_{i+m-1}^x\
\label{e:dualpauliz}
\end{eqnarray}
in terms of which the Hamiltonian in Eq.~(\ref{e:hamilton}) is expressed as
\begin{equation}
{\cal H}=-
J \sum_l \tau_l^z-h\sum_l \tau_l^x \tau_{l+1}^x\dots \tau_{l+m-1}^x\;.
\label{e:dualham}
\end{equation}
Consequently the two sets of energies $E(J,h)$ and $E(h,J)$ 
are equal:
\begin{equation}
E(J,h) = E(h,J)  
\label{e:dual}
\end{equation}
and the self-dual point $h=J$ corresponds
to the transition point of the system, provided there is one single
phase transition in the system. The duality symmetry, as described above
holds in the thermodynamic limit, i.e. when the length of the system
$L \to \infty$. 
In a {\em finite system} duality generally relates sectors
of the Hamiltonian with different boundary conditions.
With periodic boundary conditions one has the symmetries $\sigma^x_{L+i}=
\sigma^x_i$ and $\sigma^z_{L+1}=\sigma^z_i$, which in terms of the
dual operators in Eq.~(\ref{e:dualpaulix}) 
and Eq.~(\ref{e:dualpauliz}) are only satisfied in the
ground state sector of the Hamiltonian. As a result self-duality
holds only in the ground state sector, which is indeed verified
numerically.
Based on this observation we expect somewhat
different statistical properties of the energy 
levels in the self-dual and non-selfdual sectors.

\section{Results of the random matrix theory}

Using the symmetries as described in the previous Section we have
performed the block-diagonalization of the eigenvalue matrices for
large finite lattices, the size of which was a multiple of the
length of the interaction $m$. We went up to $L=18,16$ and 15 for
$m=3$, $m=4$ and $m=5$, respectively. The size of the blocks, as seen
in Tables I-III, is
relatively small, especially for larger values
of $m$ their size is reduced by gauge symmetry.

Having the block-diagonalized Hamiltonian we solved their spectrum
by standard numerical methods, which are contained in the LAPACK library.
The next step, before performing the analysis, is to unfold the spectrum,
i.e. to subtract the average tendency and to keep only the fluctuations,
which are normalized in the same manner at each part of the spectrum.
Technical details relating to unfolding the spectrum are given in
Ref\onlinecite{jc2,jc3}.

The unfolded spectrum is then analyzed and several spectral quantities
are determined and compared with the predictions of random matrix theory.
First, we consider the level spacing distribution, $P(s)$, 
which is expected
of the Wigner form in Eq.~(\ref{e:wigner}) 
for non-integrable models, whereas
it is generally of the Poissonian form for integrable models. To analyze
realistic spectra it is often useful to consider Brody's interpolation
formula:
\begin{equation}
P_{\beta}(s)=c(1+\beta)s^{\beta} \exp\left(-cs^{\beta+1} \right)\;,
\label{e:brody}
\end{equation}
with 
$c=\left[ \Gamma \left({\beta+2 \over \beta+1} \right) \right]^{1+\beta}$,
which corresponds to the Wigner and the Poisson form for $\beta=1$ and
$\beta=0$, respectively. The interpolation parameter $\beta$, 
which is determined by an optimalization fit, proved itself to
be a useful indicator for the localization of integrable 
varieties\cite{jc2,jc3}.

Another quantity characterizing the independence of the eigenvalues
is the spectral rigidity in an interval of length $l$:
\begin{equation}
\Delta_3(l)=\left\langle {1 \over l} \min_{a,b} \int_{\alpha-l/2}^
{\alpha+l/2}(N_u(\epsilon)-a\epsilon-b)^2 d\epsilon \right\rangle_{\alpha}\;,
\label{e:rigidity}
\end{equation}
where $N_u(\epsilon) \equiv \sum_i \Theta(\epsilon-\epsilon_i)$ is the
integrated density of unfolded eigenvalues and $\langle \dots \rangle_{\alpha}$
denotes an average over $\alpha$. Finally, we shall also consider the number
variance $\Sigma^2(l)$ defined as the variance of the number of unfolded
eigenvalues in an interval of length $l$:
\begin{equation}
\Sigma^2(l)=\left\langle \left[N_u(\epsilon + l/2)-N_u(\epsilon - l/2)-l
\right]^2 \right\rangle_{\epsilon}\;,
\label{e:variance}
\end{equation}
where the brackets denote an averaging over $\epsilon$.

\begin{figure}
\caption{Level spacing distribution for L=15 M=3 h/J=1.36 and ALL the 
representations. The exponential ("E") and the Wigner ("W") distributions are shown
(full line),
together with the Brody distribution (broken line)
for the fitted  best value of the parameter $\beta=1.01$ (see text).}
\label{f:pdsgeneric}
\end{figure}

First we presents the results of the statistical analysis of the spectra
{\it outside the transition point}. As seen in 
Figs.~\ref{f:pdsgeneric} and \ref{f:S2D3generic} on the example of
the $m=3$ model at a coupling $J/h=1.36$ for a
15-site chain, all the three characteristic
quantities of the spectrum are very well described by the Wigner
distribution \cite{wigner}. 
Fig.~\ref{f:S2D3generic} presents the rigidity
and the number variance for the same parameters.
The expected behavior for independent random energies
and for the eigenvalues of the GOE are also shown.
The GOE behavior is observed up to quite
large values of $L$, indicating that GOE matrices
provide a good description of the Hamiltonian.
The data shown are 
obtained averaging over
all the representations. However very similar results
are obtained averaging only over the self-dual sector.
The characteristic
parameters of the spectrum do not depend on whether the sector under
consideration is selfdual or non-selfdual
We also note that very similar behavior is found for
other ranges of the interaction $m>3$ 
or for other values of $J/h>1$.
Consequently there is no
evidence for the integrability of the model with $m>2$ away from the
critical point.

\begin{figure}
\caption{Rigidity (a) and variance (b) 
for L=15 M=3 h/J=1.36 and ALL the representations.
The corresponding quantities for the GOE matrices and for
random diagonal matrices are also shown in full lines.}
\label{f:S2D3generic}
\end{figure}

In the following we investigate the level statistics of the model
as a function of the ratio
$h/J$ and calculate the interpolation parameter $\beta$ in
Eq.~(\ref{e:brody}) as a best fit over the self-dual and
non-selfdual sectors. The results are shown in Fig.~\ref{f:betadeh} 
for $L=15$ and $m=3$, whereas data for the largest system size
are only included
in the self-dual sector. We note that the corresponding
data for $m=4$, $m=5$, and a less extensive
calculation for $m=6$, lead us to very similar conclusions.
One can see in Fig.~\ref{f:betadeh} that Brody's parameter $\beta$
has different behavior in the self-dual and in the
non-selfdual sectors.
While in the non-selfdual sectors $\beta$ is approximately
constant and its value $\beta \approx 1$ corresponds to the GOE
result, in the self-dual sector there is a change in the value
of $\beta$ around the self-dual point. Actually its value drops
from $\beta=1$ to about $\beta \approx 0.45$ at the transition
point.
The region, where the change in
$\beta$ takes place seems to shrink only to the self-dual point
in the thermodynamic limit, as can be seen in Fig.~\ref{f:betadeh}
by comparing the results with $L=15$ and $L=18$.

\begin{figure}
\caption{Parameter $\beta$ as a function of $h/J$ for $m=3$.
For $L=15$ the data are averaged separately over all 
self-dual sectors or over all non self-dual sectors. For $L=18$ only
the representation $R_0$ to which the ground state 
belongs is taken into account.}
\label{f:betadeh}
\end{figure}

This observation leads us to study 
carefully the spectral properties of the models {\em at the
transition point}, the results of which are shown in
Fig.~\ref{f:mixture}(a).
As seen in the figure the level spacing 
distribution could not be
well fitted by the interpolation formula in Eq.~(\ref{e:brody}), 
at least with the symmetries we have taken into 
account. However, it 
is given approximately by the arithmetic average of the Wigner and Poisson
distributions, which
is also shown in the figure. We argue that the measured spectral quantities
in the self-dual sector can be interpreted as if the spectrum is composed of
two independent Wigner distributions. To check our assumption we have
taken two non-selfdual blocks of roughly the same size each of which
has Wigner characteristics and merged the levels of the two blocks.
Then we analyzed the level statistics of the combined blocks and the
obtained results in Fig.~\ref{f:mixture}(b) looks very similar to those
we found for the self-dual sector 
and presented in Fig.~\ref{f:mixture}(a).

\begin{figure}
\caption{a) Level spacing distribution for L=18 M=3 and for
the representation to which the ground-sate belongs.\\
b) Combination of the spectrum of two different representation $R_{15}$
and $R_{16}$ of the $m=5$ model for $L=15$.}
\label{f:mixture}
\end{figure}

Thus at this point we conclude that the spectrum of the self-dual sector
at the self-dual point is seemingly composed of two independent parts,
each having Wigner-type characteristics. This type of behavior is the
result of an extra symmetry, the {\it self-duality}, which is just seen
in the selfdual sector. Furthermore, we argue that this extra symmetry
is manifested by the crossing of energy levels at the self-dual point in
the thermodynamic limit $L \to \infty$. To see this we have calculated the
quantity:
\begin{equation}
\delta= 
\left \langle \max(s_i,s_{i-1}) \over \min(s_i,s_{i-1})\right\rangle_i\;,
\label{e:crossing}
\end{equation}
which measures the asymmetry in the the level spacing distribution. 
For {\em independent} random variables chosen
according to a Wigner distribution one has:
$P(\delta) = \frac{4 \delta}{(1+\delta^2)^2}$
yielding a mean value $< \delta > = 1 + \pi/2 \approx 2.5708$.
For matrices from the GOE, the correlation
between spacings $s_i$ modifies this value. 
We have numerically found 
\footnote{diagonalizing 1000 GOE random matrices of size
ranging from 3 up to 2000.}
that this value does not
vary considerably with the size of the matrix, and
is very close to 
\begin{equation}
< \delta >_{\rm GOE} \approx 3
\label{e:dGOE} 
\end{equation}
As seen in Fig.~\ref{f:delta}
for non-selfdual sectors $\delta$ is indeed close to the GOE result in
(\ref{e:dGOE}). Similarly, $\delta \approx 3$ is found in the self-dual
sector far from the self-dual point, however there is a sharp increase in
$\delta$ in the neighborhood of $h/J=1$.\cite{analdelta} 
Since the value of
$\delta$ at the self-dual point is monotonically increasing with the size
of the system (see Tab.~\ref{t:delta}), 
one expects that in the thermodynamic limit $\delta \to \infty$.
Thus there is an exact degeneracy in the selfdual sector at the
transition point, which should be connected with the presence of an
extra symmetry. The possible origin of this extra symmetry is discussed
in the final Section.

\begin{figure}
\caption{The asymmetry parameter $\delta$ 
in Eq.~(\ref{e:crossing}) as a function of $h/J$
for $L=16$ and $m=4$. The upper curve corresponds to the self-dual 
sector and the lower curve to non self-dual sectors.
The error bars are calculated as the variance 
divided by the square root of the number of spacing ratios.}
\label{f:delta}
\end{figure}

\section{Level statistics at a first-order transition point}

In this paper we have studied the statistical properties of the spectrum
of a transverse-field Ising spin chain with $m$-spin interactions and
compared to the predictions of random matrix theory. Away from the
transition point, which is known exactly from duality symmetry, the
spectrum is shown to be a Gaussian Orthogonal Ensemble and its
properties are well described by the Wigner distribution. On the
other hand at the transition point the spectral properties of the
selfdual and non-selfdual sectors are different. While the spectra of
non-selfdual sectors are close to the Wigner distribution the same for
the selfdual sector can be described as the composition of two
independent Wigner distributions. Furthermore, we have shown that
this special behavior at the transition point is the result of
level crossing in the thermodynamic limit.

This observation can be compared with the known
exact\cite{kotecky} and numerical\cite{igloisolyom}
results on the two-dimensional $Q>4$ Potts model. 
As known exactly\cite{baxter},
this model is also self-dual and there is a first order transition
in the system. As an analogous quantity to the 
Hamiltonian in (Eq.~\ref{e:hamilton}) we
consider the ${\cal T}$ transfer matrix of the Potts model, which in the
Hamiltonian limit\cite{kogut,solyom} is given by ${\cal T}=\exp(-\tau H_P)$,
where $\tau$ denotes the lattice spacing and $H_P$ is the Hamiltonian
of the one-dimensional quantum Potts model. According to exact 
results\cite{kotecky}
in the thermodynamic limit the ground state of $H_P$ at the transition
point is $(Q+1)$-fold degenerate. At this point the first two levels of the
self-dual sector, as well as the first levels of the $Q-1$ other, non- selfdual
sectors are degenerate. Thus the first order nature of the transition is
manifested by a level crossing in the self-dual sector. (For finite rings
one observes a hybridization gap in the selfdual sector, which vanishes
exponentially with the size of the system\cite{igloisolyom}.) As shown by
numerical calculations\cite{igloisolyom} the same type of level crossing
phenomena takes place for the higher lying levels, too. Thus, for finite
systems, the spectrum
at the transition point is expected to decompose into two parts, which are
going to be degenerate in the thermodynamic limit.

Our numerical results on the multispin coupling model are in agreement with
the above picture, thus we expect a similar scenario. The selfdual symmetry
at the transition point, which is connected to a level crossing in the
selfdual sector in the thermodynamic limit is responsible for the unusual
spectral properties of the multispin coupling models for $m>3$. The $m=3$
model, in which the transition is expected to be second order, is assumed
to represent the border limit of continuous models. Thus one expects
that the above scenario, which stays valid as 
$m \to 3^+$, could hold also for
$m=3$, perhaps with another type of functional form for the size dependence
of the hybridization gap. Indeed, this type of behavior has been found by our
numerical studies.

Finally, we turn to discuss possible integrability of the multispin
coupling model in Eq.~(\ref{e:hamilton}). As a result of our numerical
studies of spectral properties of the model we conclude that there is
no evidence in favor of integrability of the Hamiltonian in 
(Eq.~\ref{e:hamilton}) for $m>2$, even at the transition point.

\acknowledgments
This study were partially performed during our visits in
Budapest and Grenoble, respectively, which was subsidized
by an exchange grant of the R\'eseau Formation-Recherches France-Hongrie 
(Minist\`ere de l'Education Sup\'erieure et de la Recherche).
 F.\ I.'s work has been supported by the Hungarian
National Research Fund under grants No OTKA TO12830 and OTKA TO23642
and by the Minister of Education under grant No. 0765/1997.

%
\begin{table}
%
%
\begin{tabular}{|c||c|c|c|c|c|c|c|c|c|c|c|c|c|c|c|c|c|c|}\hline
\multicolumn{19}{|c|}{M=3 L=18} \\ \hline
label R&
0& 1& 2& 3& 4& 5& 6& 7& 8& 9& 10& 11& 12& 13& 14& 15& 16& 17 \\ \hline
dimension&
1& 1& 1& 1& 2& 2& 2& 2& 2& 2& 2& 2& 3& 3& 3& 3& 6& 6 \\ \hline
size&
2029& 1871& 1645& 1743& 3613& 3671& 3612& 3668& 3668& 
3612& 3612& 3668& 5656& 5272& 5400& 5528& 10920& 10920  \\ \hline
\end{tabular}
\caption{Dimensions of the irreducible representations (i.e. degeneracy) and 
size of the corresponding block.}
\label{t:m3}
\end{table}
%
%
\begin{table}
\begin{tabular}{|c||c|c|c|c|c|c|c|c|c|c|c|c|c|c|c|c|c|}\hline
\multicolumn{18}{|c|}{M=4 L=16} \\ \hline
label R&
0& 1& 2& 3& 4& 5& 6& 7& 8& 9& 10& 11& 12& 13& 14& 15& 16 \\ \hline
dimension&
1& 1& 1& 1& 1& 1& 1& 1& 2& 2& 2& 2& 2& 2& 2& 2& 2 \\ \hline
size&
330& 265& 202& 265& 288& 288& 224& 224& 529& 512& 
544& 480& 544& 480& 496& 512& 526 \\ \hline \hline
label R&
17& 18& 19& 20& 21& 22& 23& 24& 25& 26& 27& 28& 29& 
30& 31& 32& 33 \\ \hline
dimension&
2& 2& 2& 2& 2& 2& 2& 2& 2& 4& 4& 4& 4& 4& 4& 4& 8 \\ \hline
size&
512& 496& 512& 496& 512& 526& 512& 496& 512& 1088& 960& 
1024& 1024& 1024& 1024& 1024& 2048 \\ \hline 
\end{tabular}
\caption{Dimensions of the irreducible representations (i.e. degeneracy) and 
size of the corresponding block.}
\label{t:m4}
\end{table}
%
%
\begin{table}
\begin{tabular}{|c||c|c|c|c|c|c|c|c|c|c|c|c|c|c|c|c|c|c|}\hline
\multicolumn{19}{|c|}{M=5 L=15} \\ \hline
label R&
0&1&2&3&4&5&6&7&8&9&10&11&12&13&14&15&16&17 \\ \hline
dimension&
1&1&2&2&2&2&2&2&2&5&5&5&5&5&5&10&10&10\\ \hline
size&
102&38&138&136&136&136&136&136&136&374&310&374&310&374&310&682&682&682
 \\ \hline
\end{tabular}
\caption{Dimensions of the irreducible representations (i.e. degeneracy) and 
size of the corresponding block.}
\label{t:m5}
\end{table}

\begin{table}
\begin{tabular}{| c  || c |c | c | c | c  |}\hline
                &L=10& L=12 & L=15 & L=16 &L=18 \\ \hline
             m=3&    & 4.79 & 6.54 (9.63) &  &  11.31   \\ \hline
             m=4&    & 4.58 &  & 5.09 (8.67)  &     \\ \hline
             m=5& 1.76 &     & 4.33 &   &      \\  \hline
\end{tabular}
\caption{The asymmetry parameter $\delta$ in Eq.(\ref{e:crossing})
as a function of $L$
for $m=3,4$ and $5$  in the ground-state sector.
When present the number in parenthesis refers to the asymmetry
calculated for the {\em entire} self-dual sector.}
\label{t:delta}
\end{table}


\begin{references}
\bibitem{analdelta}
We have observed sharp peaks in $\delta$ at some specific values of $h/J$
as a result of almost degenerate energy levels. We have got rid
of this finite size effect by discarding the 
values of $\delta_i$ which are 100 times larger than the average
value. 
This does not change the value of the asymmetry $\delta$
for the self-dual point $h/J=1$ since the increase at this point
is not due to some-quasi degeneracy, but to the
fact that almost all eigenvalues are clustered in pairs of close values.
\end{references}
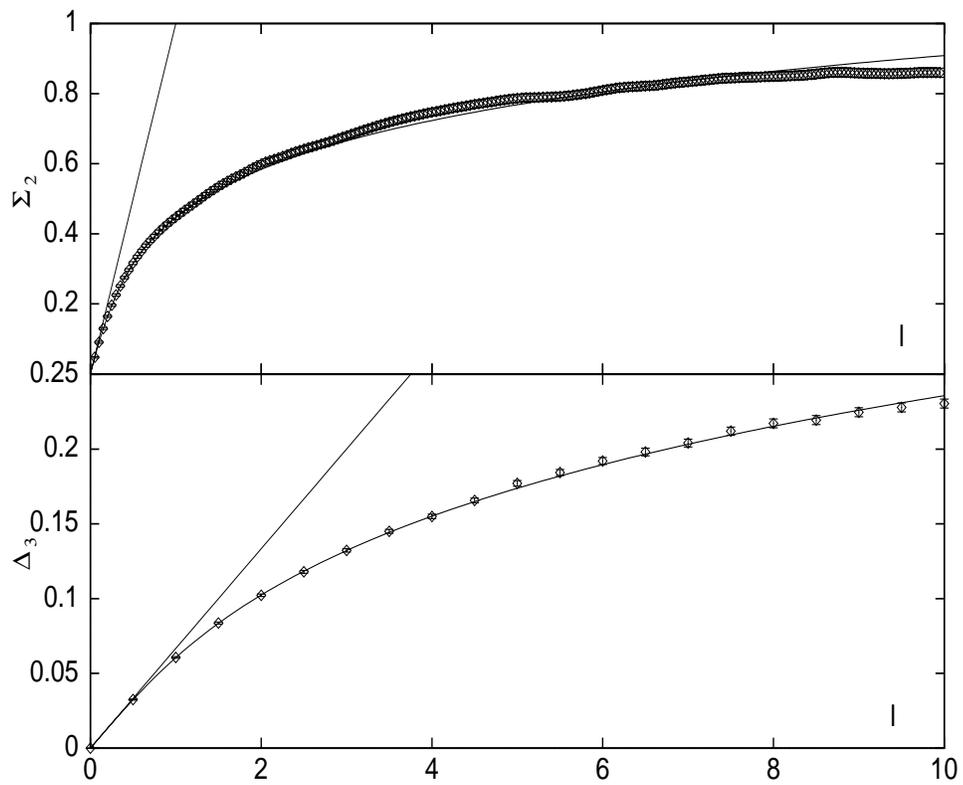
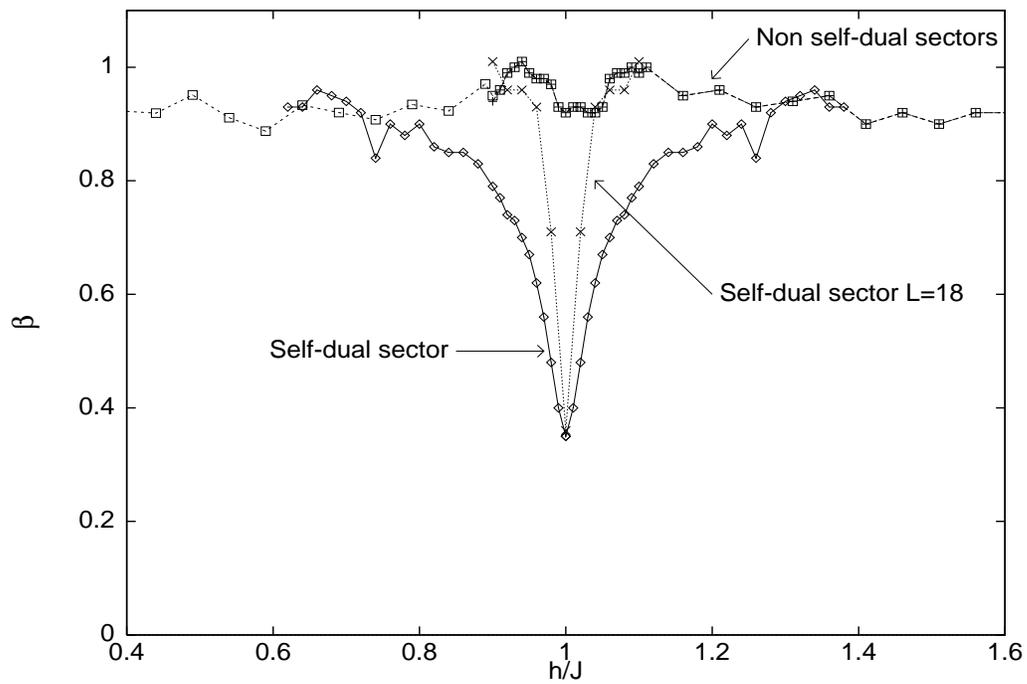
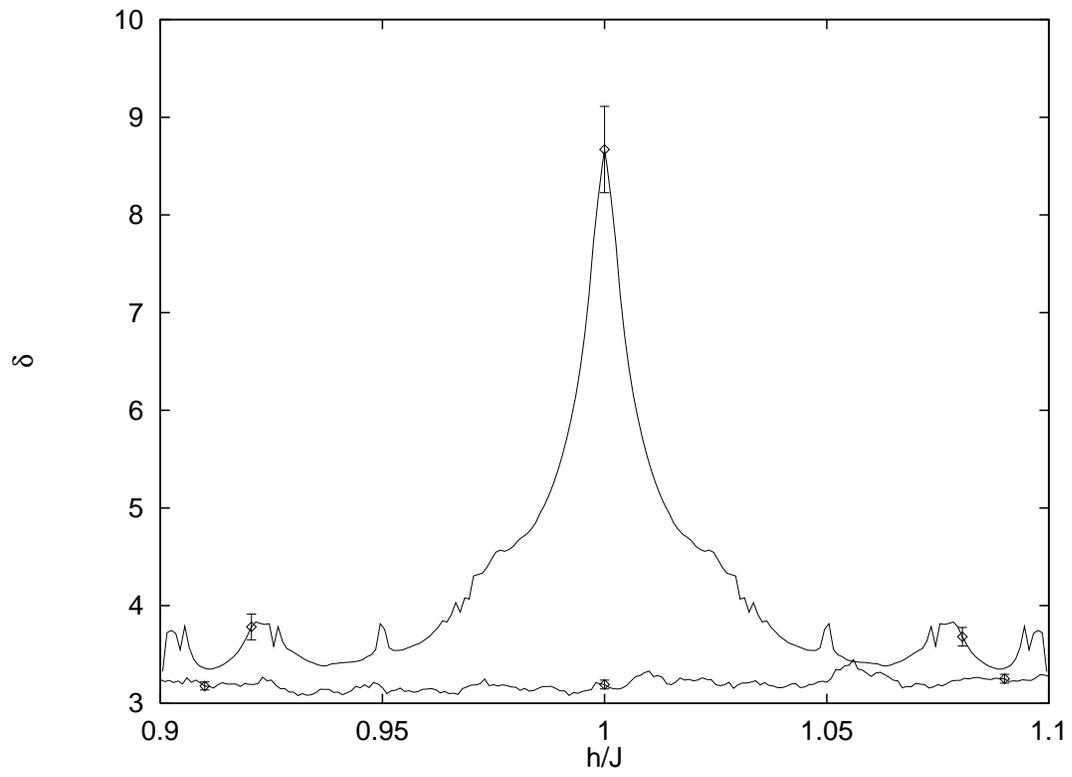
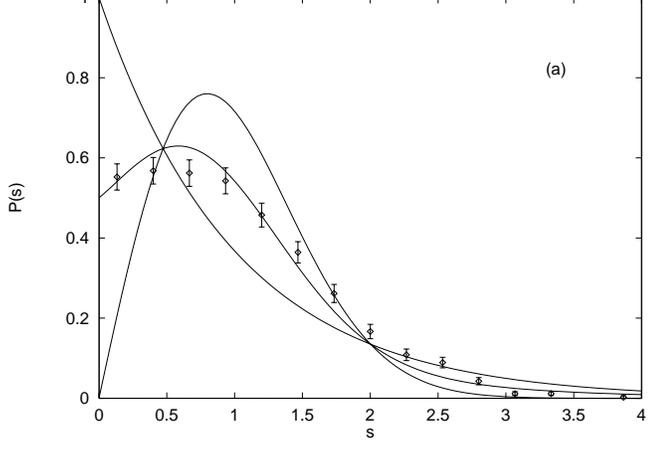
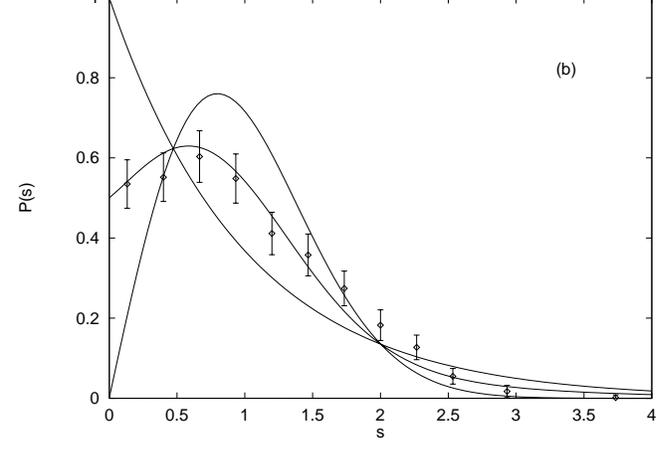
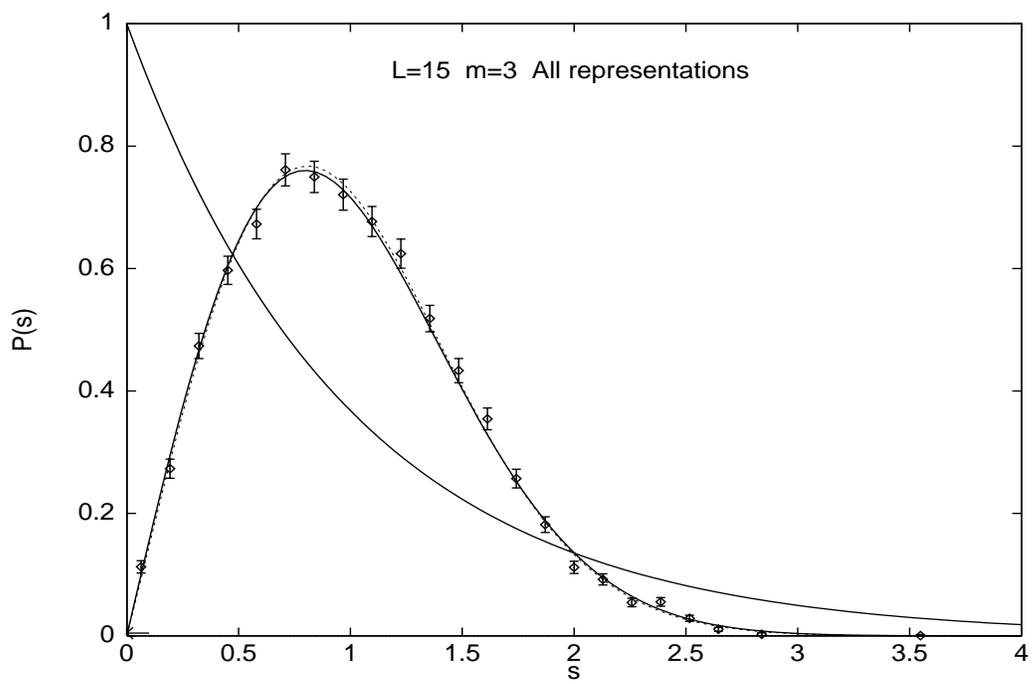
\end{document}